\documentclass{article}

\usepackage{microtype}
\usepackage{graphicx}
\usepackage{subfigure}
\usepackage{booktabs} 
\usepackage{textcomp}
\usepackage{listings}
\usepackage{hyperref}

\usepackage[accepted]{sysml2019}

\sysmltitlerunning{ISAM: A Compute and Hardware Agnostic Deep Learning Compiler}

\begin{document}

\twocolumn[
\sysmltitle{ISA Mapper: A Compute  and Hardware Agnostic Deep Learning Compiler}

\begin{sysmlauthorlist}
\sysmlauthor{Matthew Sotoudeh}{ucd,intel}
\sysmlauthor{Anand Venkat}{intel}
\sysmlauthor{Michael Anderson}{intel}
\sysmlauthor{Evangelos Georganas}{intel}
\sysmlauthor{Alexander Heinecke}{intel}
\sysmlauthor{Jason Knight}{intel}
\end{sysmlauthorlist}

\sysmlaffiliation{ucd}{Department of Computer Science, University of California, Davis}
\sysmlaffiliation{intel}{Intel Corporation}
\sysmlcorrespondingauthor{Jason Knight}{jason.knight@intel.com}

\sysmlkeywords{Machine Learning, SysML}

\vskip 0.3in

\begin{abstract}
  
    Domain specific accelerators present new challenges and opportunities for
    code generation onto novel instruction sets, communication fabrics, and
    memory architectures.

    In this paper we introduce an intermediate representation (IR) which
    enables both deep learning computational kernels and hardware capabilities
    to be described in the same IR. We then formulate and apply
    \textit{instruction mapping} to determine the possible ways a computation
    can be performed on a hardware system. Next, our \textit{scheduler} chooses
    a specific mapping and determines the data movement and computation order.
    In order to manage the large search space of mappings and schedules, we
    developed a flexible framework that allows heuristics, cost models, and
    potentially machine learning to facilitate this search problem.
    
    With this system, we demonstrate the automated extraction of matrix
    multiplication kernels out of recent deep learning kernels such as
    depthwise-separable convolution. In addition, we demonstrate two to five
    times better performance on DeepBench sized GEMMs and GRU RNN execution
    when compared to state-of-the-art (SOTA) implementations on new hardware
    and up to 85\% of the performance for SOTA implementations on existing
    hardware.

\end{abstract} ]

\printAffiliationsAndNotice{}  

\section{Introduction}

Modern computer programs are typically written in high-level programming
languages that abstract away details of individual hardware architectures. To
that end, a large body of work exists in the field of compilation techniques,
the process of automatically translating high-level program descriptions into
the low-level instruction set understood by the hardware. Crucially, there are
usually many (infinite) mappings from high-level program to low-level
executable, and the compiler is charged with finding a close to optimal (with
respect to program size, execution time, or energy use) low-level executable
that preserves the computational semantics of the high-level program.

Historically, existing work has focused on general-purpose compilers such as
GCC and LLVM that compile general-purpose input
programs written in high-level languages like C to a compute device following
the traditional Harvard or Von Neumann architectures composed of caches in a memory hierarchy
and single CPU operating on scalar or vector values.

Unfortunately, in domains such as dense linear algebra, despite decades of
compiler work it is still widely accepted that hand written and optimized
assembly surpasses the performance of code output by today's standard
compilers. Additionally, the demand for performance in these domains is driving
further hardware innovation which only exacerbates these existing problems in this domain.

Some of the issues exhibited by general-purpose compilers such as GCC or LLVM are:
\begin{enumerate}
    \item Such compilers assume that the code is being compiled for a single,
    synchronous compute unit or multiple devices with particular forms of
    parallelism and shared memory capabilities.
    \item They assume a particular form of memory hierarchy, with 
    a large main memory accessible by the CPU and a cache hierarchy on the chip
    that is managed completely by hardware.
    \item They assume a scalar or vector instruction set, and are unable to map
    computations programs onto broader types of instructions like matrix multiplication.
\end{enumerate}

In response to these issues, a number of \textit{domain-specific} deep learning
compilers have been proposed to address various aspects of the problem.


In particular, TVM~\cite{chen2018tvm} builds on the work of
Halide~\cite{ragan2013halide} to allow users to express computational kernels
in a high level description language (essentially a ``Tensor IR") and then
expose a device specific set of scheduling primitives for users to describe
loop blocking, memory prefetching, and other considerations to lower the
computational description to efficient code in LLVM intermediate representation
(IR) or other representations. The requirement that users manually schedule
kernels was partially addressed in the AutoTVM~\cite{autotvm} extension by
leveraging learned cost models, but this still requires good initial schedules.
In addition, current support for \textit{pattern matching} larger blocks of
compute such as a matrix multiplication from the finer grained IR of TVM is
inflexible given the memory hierarchy and instruction set assumptions made in
TVM.

PlaidML~\cite{plaidml} has a similar ``Tensor IR" called Tile, but uses
parameterized cost models and careful memory hierarchy analysis to generate
automatically-scheduled kernels. But again, both platforms still make strong
assumptions about the underlying memory hierarchy and instruction sets
supported by the individual compute units.

The Intel\textregistered{} nGraph\texttrademark~\cite{cyphers2018intel} library and
TensorFlow\texttrademark\footnote{Other names and brands may be claimed as
the property of others.}'s XLA~\cite{tfxla} compiler both take as input a higher
level, coarser, graph based representation of a deep learning computation and
then allow each hardware backend to choose how to lower the coarse grained deep
learning computations to machine code.  As an example, nGraph\texttrademark's CPU backend
leverages the Intel\textregistered{} Math Kernel Library for Deep Neural
Networks (Intel\textregistered{} MKL-DNN) and Eigen for much of the execution
capabilities whereas the XLA CPU backend lowers operators to kernel library
calls or LLVM IR. Both of these systems are capable of leveraging the
compilation approach described in this paper with the appropriate lowering pass
to the Tensor IR described in Section~\ref{sec:tensorisa}.


Less recent research has been published in the field of mapping computations
onto complex instruction set (CISC) architectures, primarily using directed
acyclic graphs to describe the computation. \cite{aho1989code,Keutzer:1987},
for example, describe both the program and the supported instructions as graphs
similar to SSA graphs used in LLVM, then perform a pattern matching step to
find isomorphisms between the two. However, such approaches in general fall
under the general-purpose language assumptions made by compilers like GCC and
LLVM, limiting their usefulness in the deep learning domain because they cannot
effectively analyze and exploit loop nests and loop-nest-reording invariances
that are extremely common in deep learning programs. Finally, they generally do
not address the problem of actually scheduling memory movement and splitting up
large computations over heterogeneous, parallel architectures.

Typically, the creation of hand written kernel libraries has been the
workaround given the limitations mentioned previously, but these libraries also
have several issues. First, the reliance on hand-written kernels means that
each new hardware architecture and instruction set requires significant
investment from the hardware vendor to even begin executing programs. Also,
whenever significant, novel kernels are introduced, even existing devices
require additions to the kernel library and/or compiler systems for support.
And finally, even when lowering rules exist, the fact that such kernels are
written and called in isolation to the rest of the program sometimes misses
optimization opportunities such as operator fusion.

\subsection{Our Approach} In this paper, we propose a compiler system to
address these issues. In particular, we limit our domain of interest to machine
learning programs such as those supported by TVM and PlaidML, and show how such
a domain-specific compiler can automatically produce optimized executable
programs for heterogeneous systems that, until now, have been unaddressed by
all existing compilation approaches.  We demonstrate the performance advantages
of our approach in a variety of cases.

We break up the compilation problem into two steps. The first,
\textit{instruction mapping}, attempts to enumerate the multitude of ways that
one program can be executed on all of the compute devices in the system. We
show how, for a limited but common set of cases, we can perform this mapping
with an automated and efficient algorithm. For more complex cases, we discuss
how we extend this approach to support arbitrary programs as well. The second
step is \textit{scheduling}, which consists of a number of choices, including:
which instructions to use, how to break up the computation, device allocation,
and memory movement throughout complex memory hierarchies. Finally, we list the
many combinatorial choices that must be made by the system and how they affect
the final executable quality, then show how we provide a unified interface to
making such choices, enabling future research to build on our system with new
cost models and heuristics.


\section{Instruction Mapping}
\label{sec:tensorisa}

One of the major assumptions modern compilers make is that the instruction set
they are compiling to (the language actually recognized by the hardware
architecture) consists primarily of scalar and vector operations. Modern
programming languages and compilers have been written so that multiple layers of IR can be 
\textit{lowered} onto predominantly scalar instruction sets. For example, in the matrix multiplication pseudo-code shown in
Listing~\ref{lst:matmul_pseudo}, the compiler could look at each line and
apply lowering ``templates" to translate each line into viable x86
instructions.

\begin{lstlisting}[caption={Pseudo-code for a matrix multiplication. For succinctness, we do not explicitly show the loop nest ordering. Traditional compilers are typically unable to analyze deep loop nests such as this one.},label=lst:matmul_pseudo,captionpos=b]
for i, j, k {
  C[i][j] += A[i][k] * B[k][j];
}
\end{lstlisting}

\begin{lstlisting}[caption={Pseudo-code describing a 1D convolution},label=lst:1dconv_pseudo,captionpos=b]
for i,x,d,k$_i$,k$_o$ {
  C[i][x][k$_o$] += A[i][x+d][k$_i$]
                    * B[d][k$_i$][k$_o$];
}
\end{lstlisting}

However, as the instructions supported by hardware become more complex,
lowering is not adequate because the granularity of the compiler IR is
often lower than instructions offered by new hardware. For example,
processing units may expose matrix multiplication instructions that can execute
thousands of multiply-accumulate operations in a single cycle. A traditional
compiler, which assumes scalar and vector instructions and works via
statement-by-statement rewriting or limited template matching, would not be able to
determine that the entire program in Listing~\ref{lst:1dconv_pseudo} can be
broken up and executed with a series of matrix multiplication instructions.

In some limited cases, a compiler might be able to support a textual template
for such matrix-multiplication programs, but this template would be limited and
not robust to syntactic changes like loop nest reording or buffer
transpositions that do not change the semantic meaning of the code.
Furthermore, we also want to be able to schedule these larger instructions
across smaller compute blocks without hardware support or explicit user
direction as is the case today with most GPU and CPU programming models.

\begin{lstlisting}[caption={Pseudo-code describing a separable depthwise convolution},label=lst:sepdepth_pseudo,captionpos=b]
for b,i,j,k,d$_i$,d$_j$,q,r {
  C[b][i][j][k] += A[b][s*i + d$_i$]
                       [s*j + d$_j$][q]
                   * D[d$_i$][d$_j$][q][r]
                   * P[c*q + r][k];
}
\end{lstlisting}

In other cases, the computation must be transformed at a coarse level of
granularity to map to fixed-function accelerator compute blocks.  For example,
Listing~\ref{lst:sepdepth_pseudo} shows pseudo-code for a separable-depthwise
convolution~\cite{sifre2014rigid}, which is a relatively recent kernel used in
the computer vision domain. This operation can be executed on
matrix-multiplication and convolution accelerators, however when expressed in
this type of IR, a number of transformations are required before the
matrix-multiplication can be pattern matched directly. We developed our IR and
compiler to be able to address these mapping and transformation challenges in
the deep learning domain.

\subsection{Representation}
We represent both the program to be executed and the instructions exposed by
the hardware in the same IR which then casts the
overarching problem as one of finding isomorphisms between sub-computations in
the ``haystack" program and the ``needle" program (describing a hardware
instruction). We first motivate and describe the IR. 

We focus our representation on a subset of programs, namely deep learning
kernels. These usually consist of simple arithmetic operations (addition,
multiplication, and subtraction) on scalar elements in high-dimensional arrays,
with indices of the arrays being determined by affine combinations on a set of
loop variables. Importantly, such kernels are almost always loop-order
invariant when ignoring floating point associativity, which is typical in the
deep learning domain. As an example, matrix multiplication is shown in
Listing~\ref{lst:matmul_pseudo}. This invariance makes such programs simpler to
analyze and optimize than general-purpose programs, as all loop reordering
operations are valid.

This property has been leveraged by deep learning compilers such as TVM and
PlaidML, which similarly restrict their input domain to such dependency-free
programs, but because they assume low-level instruction sets, these compilers
do not need to perform a mapping analysis of their programs as we describe
here.

\begin{lstlisting}[caption={ISAMIR for a matrix multiplication},captionpos=b,label=lst:matmul_isamir]
forall i,j,k {
  tmp[i][j][k] := A[i][k];
  tmp[i][j][k] *= B[k][j];
  C[i][j] += tmp[i][j][k];
}
\end{lstlisting}

To support mapping, we break up a TVM compute description
into a three-operand form, explicitly stating the individual (tensor) operations
that compose the kernel. Our intermediate representation, ``ISAMIR," is shown
in Listing~\ref{lst:matmul_isamir}. We effectively retain the iteration
order-invariance of TVM while adding the requirement that each statement
performs exactly one operation. This allows easier analysis over
sub-computations that are important in the mapping process.

Notably, for analysis purposes, our system assumes that each statement (line)
in the intermediate representation is executed in isolation over its entire
iteration domain before executing the next statement. Thus, the ``forall loop"
surrounding the program is semantically different than a for loop in languages
such as C, and here only serves to explicitly enumerate the loop axes
themselves (and not state anything about iteration order). However, this is
primarily for analysis purposes, and at execution time multiple statements can
be executed in more efficient ways, if semantic equivalence can be proven.

Furthermore, as with three-operand formats in traditional compilers, this
format requires a number of temporary buffers that would not be required in a
TVM-like description. However, these are only necessary for analysis purposes,
and will be removed or replaced before execution.

Formally, we define ISAMIR in terms of \textit{statements} over \textit{loop
domains} that act on \textit{expressions} referencing locations inside of
\textit{buffers}.

\subsection{Deterministic Mapping}
\begin{lstlisting}[caption={ISAMIR for a 1D convolution},captionpos=b,label=lst:1dconv_isamir]
forall i,x,d,k$_i$,k$_o$ {
  tmp[i][x][d][k$_i$][k$_o$] := A[i][x+d][k$_i$];
  tmp[i][x][d][k$_i$][k$_o$] *= B[d][k$_i$][k$_o$];
  C[i][x][k$_o$] += tmp[i][x][d][k$_i$][k$_o$];
}
\end{lstlisting}

Starting with an example, we can imagine mapping a program such as the
one dimensional convolution in Listing~\ref{lst:1dconv_isamir} to hardware
supporting the transposition and matrix multiplication instructions. The latter as described in ISAMIR in
Listing~\ref{lst:matmul_isamir}.

\begin{lstlisting}[caption={``Mapped" ISAMIR for a 1D convolution},captionpos=b,label=lst:mapped_1dconv_isamir]
forall i,x,d,k$_i$,k$_o$ {
  transpose(A, (1, 0, 2));
  transpose(C, (1, 0, 2));
  matmul(A[x+d][:][:], B[d][:][:],
                       C[x][:][:]);
  transpose(C, (1, 0, 2));
}
\end{lstlisting}

Listing~\ref{lst:mapped_1dconv_isamir} shows one form this mapping could take
using a modified version of ISAMIR. To see the equivalence between
Listing~\ref{lst:1dconv_isamir} and~\ref{lst:matmul_isamir}, note that the
buffers "A", "B", and "C" are used in the same way between both programs.
Furthermore, loop axes in the original program have corresponding axes in the
mapped instructions, for example, k$_i$ is used exactly as $k$; as the minor
and major dimensions on the right hand side of the first two statements, then
it is summed over in the final statement. Note that there were multiple choices
for mapping the $k$ axis, we only show one here.

In the general case, we wish to determine a \textit{buffer map},
\textit{dimension map}, and \textit{axis map} between the haystack and needle
programs, after which we can rearrange the dimensions and axis such that the
mapped dimensions and axis are inner-most in the computation and can be
replaced by a single instruction call. For a well-defined subset of cases (such
as single-reductions into output buffers) such mappings preserve semantic
equivalence with the input program.

To perform this mapping process in general, we represent each buffer access
expression as a matrix, with buffer dimensions on the rows and loop axis on the
columns. Now we can express buffer and axis maps as \textit{permuted subsets}
of the matrix rows and columns, respectively. The goal now becomes to select
rows and columns, along with their orderings, from the haystack matrices such
that the selected sub-matrix is equivalent to the one exposed by the target
instruction.

Effectively, and similar to the representations used in many polyhedral
compilation techniques, these matrices map from the values of the loop axis at
a particular iteration to a specific element in the buffers to be acted upon.
When combined with the domains of the loop axis and the fundamental operation
to be executed (such as ``*=" and ``+="), these matrices fully describe the
computation that is being performed. Thus, mapping equivalent sub-portions of
these matrices intuitively maps equivalent sub-computations between two
programs.

Notably, by treating these matrices as graphs, our problem is isomorphic to the
bipartite subgraph isomorphism problem which has complexity no greater than the general
subgraph isomorphism problem $\mathcal{O}(n^{0.729w})$ where $n$ and $w$ are the number of
vertices of the graph representation of the source program and target hardware
instruction respectively~\cite{nevsetril2012sparsity}.

Taking inspiration from existing subgraph isomorphism solvers~\cite{sgisolver},
we devise a dynamic programming algorithm that recursively attempts to map
buffer dimensions and axes until all possibilities are exhausted
or a mapping does not hold. Notably, the recursive nature of this mapping
means that entire ``branches" of possible mappings can be ignored as soon as
one of the mappings is shown not to hold. We utilize a number of heuristics to
further speed up the algorithm, and its execution time was negligible for all tested
programs.

\subsection{IR Transformations}
Although the deterministic mapper is able to automatically and efficiently
determine isomorphisms in many real-world programs, there are classes of
programs where matching is insufficient alone. For example, in
separable-depthwise convolutions shown in Listing~\ref{lst:sepdepth_pseudo}, an
additional multiplication is inserted into the computation. In this case, the
deterministic mapper will refuse to map the programs, as the operation order
does not match. For these cases, we must introduce additional IR
transformations for the mapping process described above to work more broadly. For example,
in the separable-convolution example, we can factor part of the second
multiplication outside of the summation, resulting in a semantically-identical
program with syntactically different statements. At this point, the
deterministic mapper can identify the (now obvious) isomorphism
between the first three instructions and a matrix multiplication, while the
last two instructions can be mapped to a dot product instruction (or, with
further transformations, a matrix multiplication).

ISAM's IR and implementation support arbitrary transformations of this form, as long as they
follow an established interface for transformation passes. Notably, these
transformations may sometimes be applied indefinitely, such as a pass that adds
new, size-1 dimensions to a tensor (to express a remove-dimension
transformation from the needle program). In addition, there is no clear order
in which to apply these passes such as to always expose the desired mapping to
the deterministic mapper. For these reasons, the non-deterministic mapper
actually creates a \textit{search space} in which the system must search for an
ideal series of transformations and mappings.

Although, on first glance, this appears to be a significant problem, we have
found in practice that the search space can be effectively managed. First, the
set of applicable transformations is very small. For a majority of operations,
no search is necessary. For situations in which search is necessary, it appears
that a small set of core algebraic transformations (expressing fundamental
algebraic facts such as associativity, commutativity, and distrubitability) are
effectively able to describe the majority of relevant transformations.

Additionally, we have found that the deterministic mapper can be used to
provide feedback to the IR transformation search. For example, in the
separable-convolution example above, the deterministic mapper can report where
and why it failed to map -- in this case, because the third operation was a
multiplication instead of an addition. The non-deterministic mapper can then
use this information, along with prior knowledge of what the factorization pass
does, to determine that performing the factorization pass would make the needed
change. For all of the use cases we have experimented with, this feedback has
been all that was needed to effectively map to matrix multiplication
instructions (obviating the need for further search techniques).

Notably, while the deterministic mapper returns axis and dimension maps, the
deterministic mapper paired with the IR transformer actually returns a set of possible mappings, each one
with an associated series of needed transformations and resulting axis- and
dimension-maps.

\begin{figure}[t!]
  \centering
    \includegraphics[width=1\linewidth]{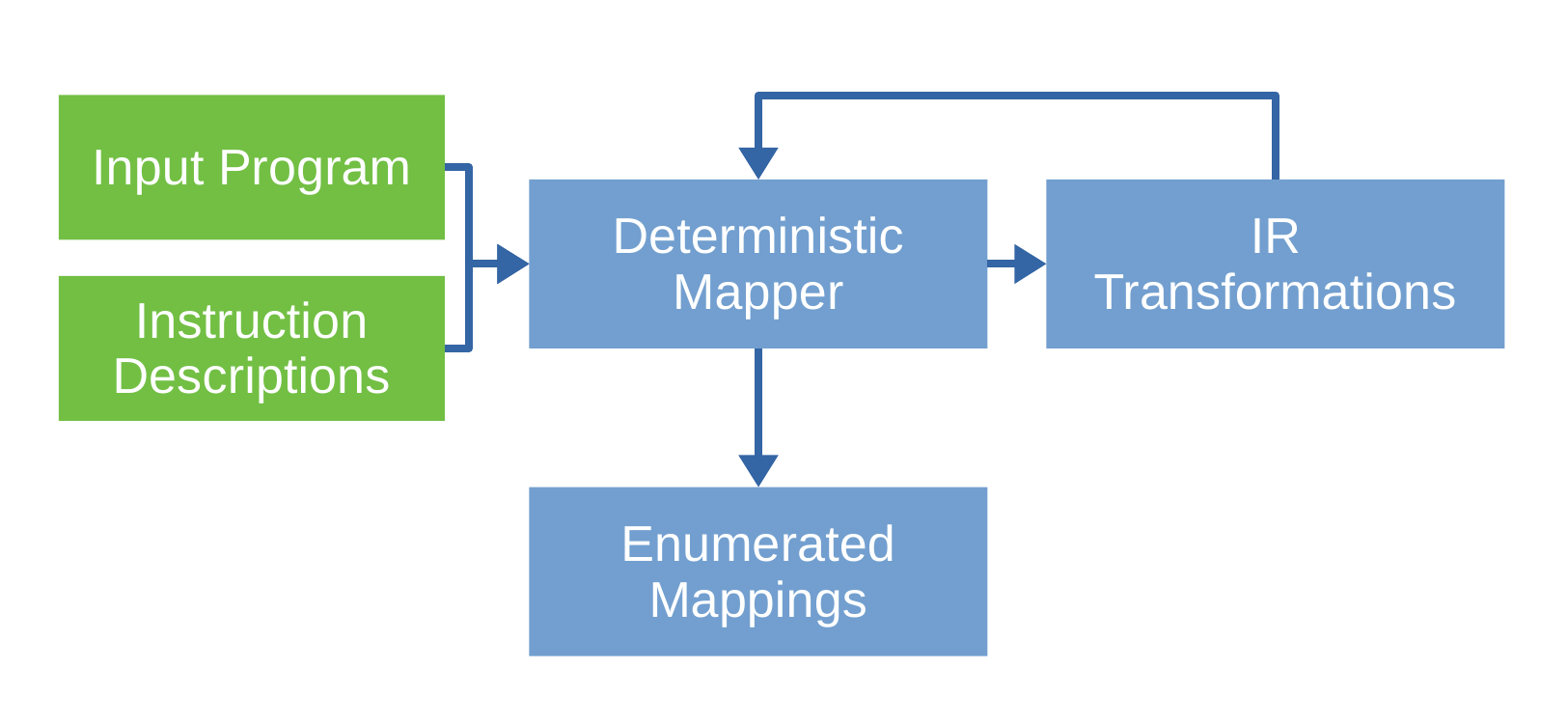}
  \caption[Mapper overview]{
      ISAM combines a fast but limited ``deterministic
      mapper" with a set of ``IR transformations" to effectively
      map complex computations onto tensor-level instruction sets.
  }
  \label{fig:mapper}
\end{figure}

Thus, the entire mapper system can be seen as a loop, shown in
Figure~\ref{fig:mapper}, where the non-deterministic mapper is constantly
sampling points from the search space, and the deterministic mapper is
analyzing each of those points for potential mappings. The result of this
process is a set of transformations, each with an associated set of axis and
dimension mappings.

\subsection{Instruction Selection}
This system often produces multiple different potential mappings for a single
input program or set of statements. For example, anything that can be mapped to
a matrix multiplication instruction could also have been mapped to a dot
product instruction. Similarly, some architectures may expose ``fused"
instructions, for which the system can choose whether to call two independent
instructions serially or the single fused instruction. We discuss our general
approach to making such compiler choices in Section~\ref{sec:approaches}, but we found that a
reasonable heuristic is to pick the non-overlapping instructions that lead to
the minimum number of final instructions used.

\section{Scheduling}
\label{sec:scheduling}

\begin{figure*}[t!]
  \centering
    \includegraphics[width=1\linewidth]{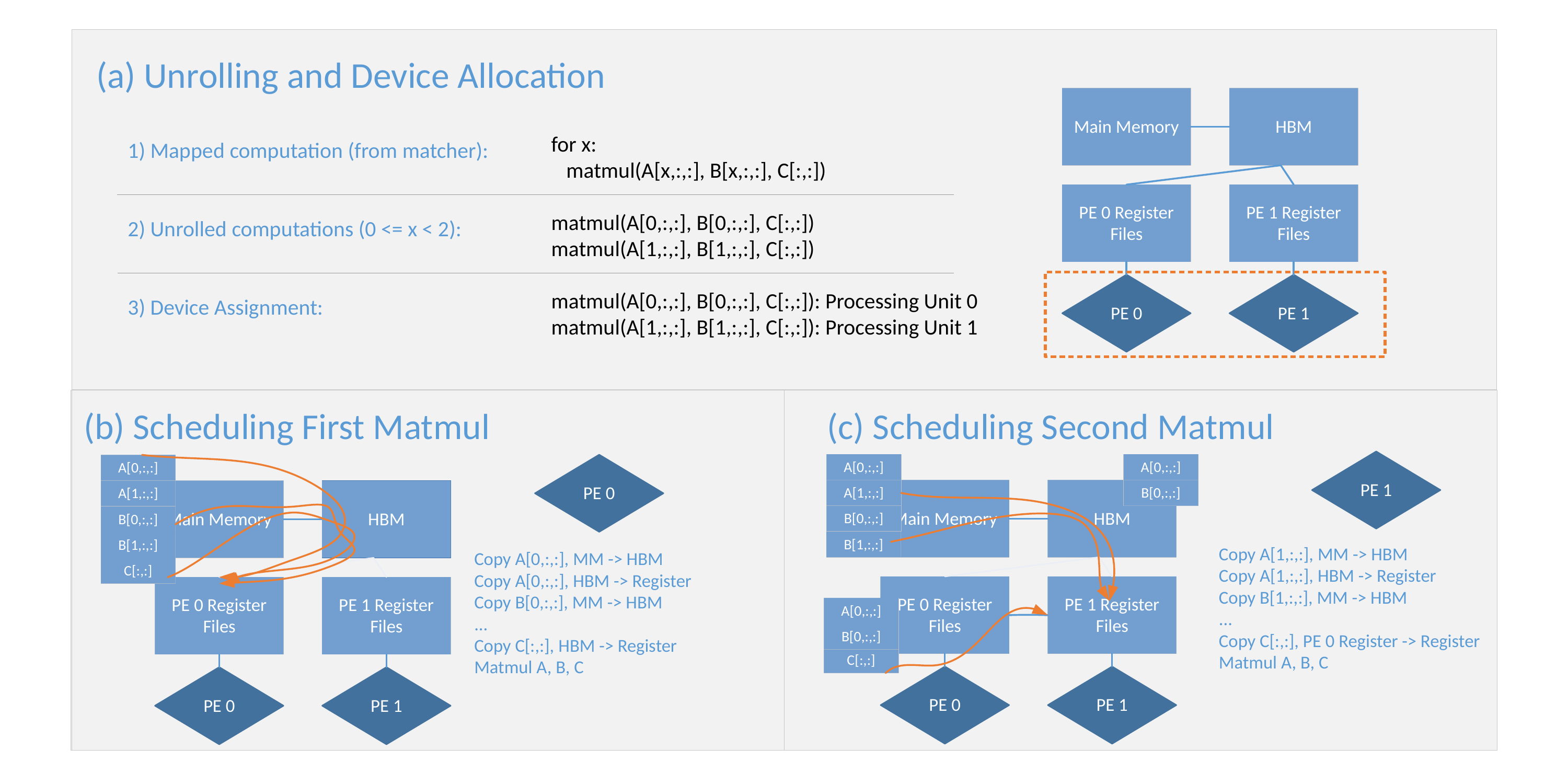}
  \caption[Scheduler overview]{
   In this small example, we attempt to execute two matrix multiplication
    instructions then sum the products together into the C matrix.  First, we
    need to determine an unrolling order and device allocation (a) for each
    individual instruction. In this example, we assume that our architecture
    has two processing units with separate registers.  After allocation and
    unrolling (b), ISAM determines how to move the appropriate memory onto the
    appropriate computation devices using graph traversals and the approach
    class described in Section~\ref{sec:approaches}, where the system records the
    necessary data movement commands and which device needs to execute them by
    labels attached to the graph hardware representation. For subsequent
    compute instructions (c), the latest version of data layout is considered
    which enables efficient data reuse amongst computing devices and registers.
    When data is updated in place, static cache invalidation is performed to
    invalidate other copies of that data across the graph.
 }

  \label{fig:scheduling}
\end{figure*}

Now that we know which instruction we wish to use to compute each portion of
the source program, the system must actually produce executable instructions that can
be executed on the target devices.

\subsection{Compile-Time Scheduling Approach}
Many existing deep learning
compilers~\cite{chen2018tvm,plaidml,tfxla,tensorcomp} make use of
\textit{static scheduling}, where the computation is broken up into
\textit{scheduling} (or compilation) and \textit{execution} phases. During the
scheduling phase, the system emits a series of instruction calls that can later
be executed to produce the desired output. This scheduling approach is unable
to adapt to changing hardware or program conditions, but produces zero runtime
overhead.

ISAM also performs static scheduling, as we have found the overhead of runtime
scheduling too great for the limited benefits. Our static scheduler operates
through a ``dry-run" approach where it attempts a simulated execution of the
program, while recording the instructions and associated system state needed to
perform the needed computation. This instruction record is then stored for final execution.

\subsection{System Description Graph}
In order to flexibly schedule across a wide variety of future 
systems, we utilize a \textit{system description graph} as an abstraction layer
for the underlying hardware. This graph is provided by the user to describe the
machine they wish to execute their program on. The system description graph
contains three types of nodes: \textit{compute nodes} that expose support for
computational instructions, \textit{memory nodes} that contain information
about available size and allocation instructions, and \textit{data movement
nodes} that describe instructions used for moving data between memory nodes
(essentially edges between the memory nodes).

Notably, these nodes are critical objects which interact during the scheduling
process by retaining the system state during scheduling.  For example, each
compute node contains the list of the instructions that it will execute at
runtime, memory nodes contain a compile-time list of memory buffers that the
system will allocate on them, and edges encode the device or devices that can
emit instructions to control memory movement across them. In this way, the
graph nodes themselves operate similarly to a hardware abstraction layer (HAL)
in a traditional compiler.

\subsection{Unrolling}
In order to analyze data dependencies more accurately, the first step of the
scheduling process is to \textit{unroll} the computation, imposing an explicit
order on the sub-computations (instruction calls) to be made. We refer to these
sub-computations as \textit{compute tiles}, and each will be associated with a
single call to the underlying instruction. In parallel computations, this
determines the \textit{dependency order} -- if two different devices need to
update the same memory, the one earlier in the unrolling will have priority and
the second will be dependent on the first's completion. We describe how this
choice is made in Section~\ref{sec:approaches}, but a reasonable heuristic is
to place computations which use the same memory close together in the
unrolling.

\subsection{Device Allocation}
In the general case, there may be multiple physical devices which can execute a
given instruction. For example, in our test architecture described later, each of the
individual compute units can execute a matrix multiplication operation. The
system must decide which compute unit will run which portion of the computation
(compute tile). We discuss how this choice is made in
Section~\ref{sec:approaches}, but a reasonable heuristic must balance memory locality with
exploitation of compute parallelism.

\subsection{Scheduling Memory Movement}
At this point through scheduling, we have assigned compute statements to devices
and their order. However, we have not expressed how the compute devices will
access the data they need to compute on. For this, we keep track of
where any given piece of data is in the system at any given stage in the
execution process.

For example, before any computation is run, we assume that all relevant memory
buffers are stored in the system memory. Now, imagine we wish to execute a
matrix multiplication instruction on a particular compute unit in our test
architecture described in Section~\ref{sec:testarch}, reading data buffers "A",
"B", and "C", then writing to buffer "C". We first ask the compute unit's
representative node in our system graph which memory units each operand for the
desired instruction may be in before execution. In this step, the compute
unit's node itself can expose limitations such as a requirement that all
operands be on different register files.

In order to place the data in one of these executable locations, the compiler
must first decide which currently-stored location it wants to copy the data
from (for example, if a copy of a network's weights are stored both on the host
memory and on the on-chip HBM), then which location it wants to copy the data
to (for example, if there are multiple register files), and finally which path
of intermediate nodes and copy instructions it wants to use to move the data
(for example, if data must be copied to an HBM unit before it can be copied to
the actual register file). All of these problems are difficult and, in many
cases, architecture-specific. We describe our general approach to such choices
in Section~\ref{sec:approaches}, however there are often good
architecture-specific ``template paths" that can be used for heuristics (for
example, copying from main memory to an HBM to a register file) and simply
finding a shortest-path tends to work relatively well. Additionally, there are
more concerns than simply the raw latency between memory units. For example,
evicting resident data from registers or HBM to make space for the desired data
may have ripple effects and can multiply the required memory bandwidth.

Furthermore, as many memory movement paths involve multiple memory units, we
keep track of intermediate copies of buffers as they are copied throughout the
system, allowing the scheduler to use these intermediate copies later as
essentially cached copies of the data. In this way, our scheduler can utilize
explicitly-allocated memory units as cache devices. However, when data is
written to a copy of a buffer in one location, our scheduler must perform a ``virtual"
\textit{cache invalidation} to note that all previous copies of the data are
now out-of-date.

\subsection{Scheduling Recurrent Models}

Recurrent models that execute groups of computations repetitively pose
additional challenges and opportunities for scheduling. For example, a GRU cell
may be executed hundreds of times in a model's execution.  Some compilers
unroll these RNN loops to reduce the problem back to standard scheduling.
However, this approach is expensive at compile-time, places further challenges
on memory optimization passes, and limits flexibility for dynamic RNN length
control.

Ideally, we would like to schedule a finite number of steps at compile-time and
invoke these sub programs dynamically at execution-time, but naively scheduling
the computation once is not sufficient since the buffers may not be in the same
location after the computation as they were in the beginning. Also, scheduling
with awareness of repeated execution offers additional optimization
opportunities such as persistent weights. To address this, ISAM explicitly
exposes the concept of a recurrent loop and schedules these loop bodies
specially three separate times. First is the \textit{priming iteration}, which
performs one instance of the computation, then leaves the data buffers as close
to the compute devices as possible. Next is the \textit{recursive iteration},
which executes on the data buffers from the priming iteration and ensures all
outputs overwrite the appropriate inputs.  Finally, the scheduler emits a
\textit{finish iteration}, which performs the computation a final time and
places the data buffers where they will be needed by the next instruction in
the program. At execution time, a driver first executes the priming iteration,
then the recursive iteration as many times as necessary, and finally executes
the finish iteration.

\section{Approaches: Unified Interface to Compiler Choices}
\label{sec:approaches}

Throughout the compilation process described in the previous sections, a number of choices have to be made that
directly affect the efficiency of the generated executable. These choices include
the strategy for interleaving deterministic mapping with IR transformations,
choosing a single instruction mapping, tiling factors, unrolling
order, device allocation, and memory movement plans.

Throughout the text we have described a few heuristics for these choices and
ISAM implements these heuristics cleanly separated from the core compiler.
But, for future work we would like to explore a number of other approaches,
including Monte Carlo tree search combined with cost modeling and/or machine
learning-guided methods. In order to facilitate this future work we developed
and employed a single interface to all such choices made by ISAM in the form of
an ``Approach" class. This decoupling of compiler infrastructure and heuristics
is similar to optimization pass interfaces in compilers such as LLVM, but
operates at multiple levels of ISAM as opposed to a single middle optimization
layer. This should enable rapid development of additional heuristics and
techniques in the compiler landscape.

In our reported results (Section~\ref{sec:results}) we utilize an Approach
class that contains a number of heuristics specific to the test architecture
described next, primarily to limit the search space of possible memory paths
and place computations using the same memory physically close to each other.

\section{Case Study Architecture}
\label{sec:testarch}

While the ISAM architecture is hardware-agnostic, we tested the principles
against both an existing CPU architecture and a novel deep learning
architecture which exhibits many of the challenges discussed previously. This
latter processor is made up of many compute units that can execute native
matrix instructions (such as matrix multiplication), element-wise operations
(useful for activation functions), and matrix-wide reductions (such as sum or
max), in addition to other special purpose units for common operations and
control flow. Furthermore, on-chip ``clusters" of these compute units are
programmed with the same instruction stream and share a set of large register
files, while there are several high-bandwidth memory modules to enable rapid
access to memory too large for the register files.  There are a number of host
and device-side instructions to move memory between register files, processing
units, and high-bandwidth units. Generally speaking, the processing units can
only execute instructions on data in their respective register files, and
further restrictions on how many times a single register file can be used in an
operation exist as well. All of the memory units in this system are explicitly
managed, so there is no cache hierarchy.

This architecture presents a challenging problem for compilers
with significant use of tensor-level instructions, large amounts of
parallelism, and a complex, explicitly-managed memory hierarchy.

This test architecture has a hand-optimized kernel library developed for it
which provides our performance baseline. For common operations and tensor
sizes, the kernels in this library are able to achieve nearly-theoretical
maximum utilization of the architecture.

In Section~\ref{sec:results} we compare cycle count comparisons between
ISAM-generated kernels and those from the existing kernel library. Due to a
number of profiling differences and issues between the kernel library and ISAM,
these numbers do not include memory movement to or from the main memory (which
should be equivalent for both approaches anyways), and ISAM kernels have some
additional overhead compared to the kernel library because of the
less-efficient way our system calls the device driver.

\section{Results}
\label{sec:results}

\subsection{Mapper}
\label{sec:results:mapper}

We first attempted to test the effectiveness of our mapping system, as described in
Section~\ref{sec:tensorisa}. To this end, we compiled a small set of
newer convolution kernels and RNN cells (including depthwise and separable
depthwise convolution and GRU cells), then used ISAM to map these kernels to a system with matrix multiplication/manipulation and element-wise
instructions. We found that, our mapper was successfully able to determine the expected mappings in all cases. Note that all previous work
requires this mapping to be explicitly specified by the programmer.

In addition, we found that, by exposing BLAS methods as ``instructions" to
ISAM, we can map the convolutions onto BLAS calls on x86 devices as well. In
Section~\ref{sec:isamtvm}, we also show how to effectively target x86 by using ISAM
mappings with TVM and LLVM. These results demonstrate how the system can be
effectively utilized even on existing devices supporting scalar and
vector instructions.

\subsection{Scheduler}

\subsubsection{GEMM}
\begin{figure}[t!]
  \centering
    \includegraphics[width=1\columnwidth]{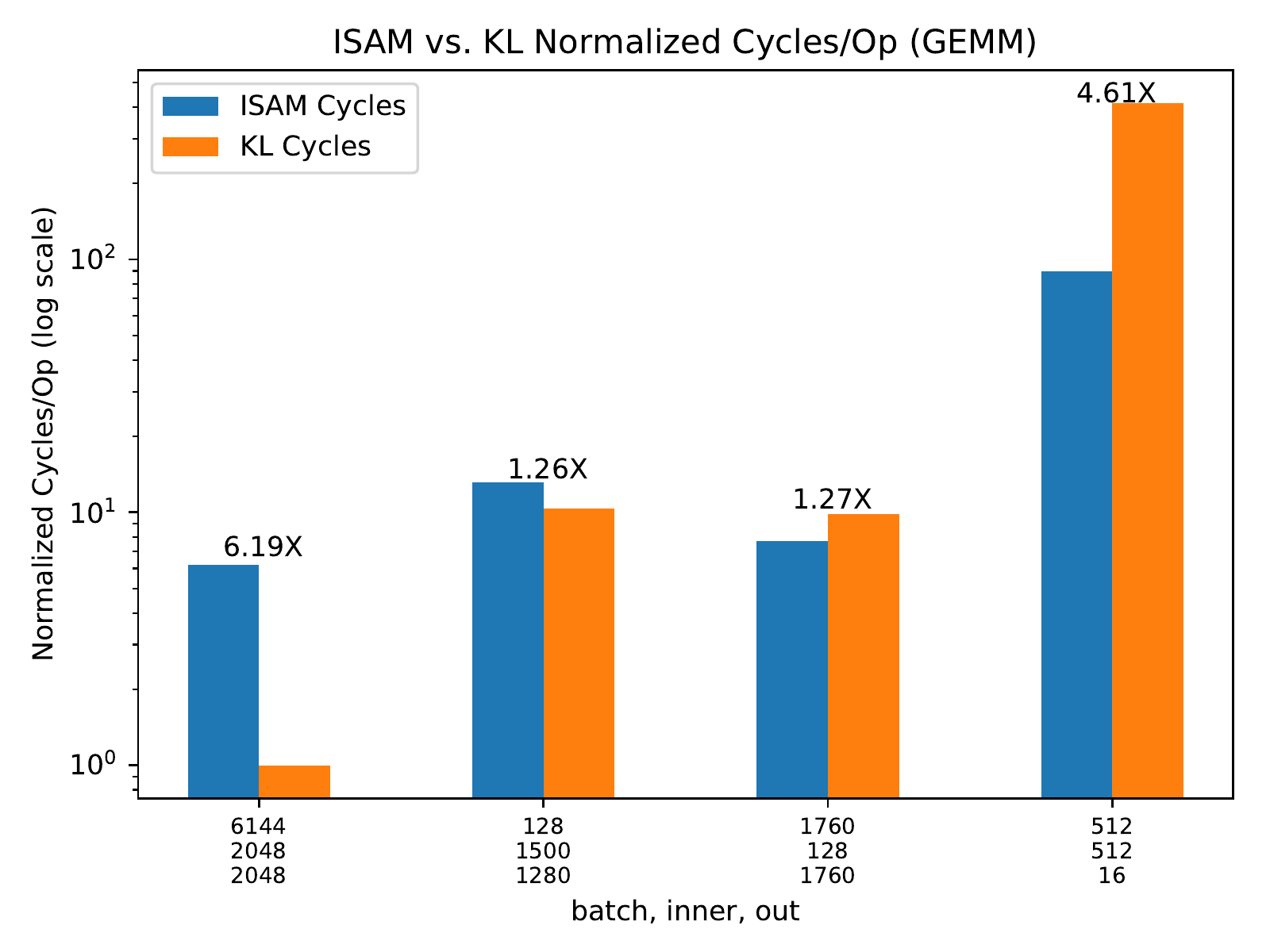}
  \caption{
      A selection of results comparing cycles per operation for ISAM and the
      kernel library (``KL") when performing a matrix multiplication. When the
      existing library is optimized for a given size, it can perform well over
      five times faster than our generated kernels.  However, for many other
      sizes that are less common or significantly different than the library's
      intended focus, ISAM can produce programs competitive with or
      significantly faster than the existing KL.  We believe that further
      heuristics and cost modeling can significantly improve ISAM's
      performance.
    }
  \label{fig:matmul_results}
\end{figure}

Next, to test the effectiveness of ISAM's scheduling we performed single matrix
multiplications with sizes from DeepBench~\cite{deepbench} and a
internal size list representing real-world usage patterns for the
architecture. We report a selection of the results in
Figure~\ref{fig:matmul_results}. For confidentiality reasons,
Figures~\ref{fig:matmul_results} through~\ref{fig:gru_results} have been
normalized by the minimum value displayed on the plot to conserve relative differences.
(thus, the ratio of any two data points is accurate, but the absolute value or
difference is not provided).

First, in
configuration (a) when the kernel library (KL) is well-optimized for a particular size,
it can significantly out-perform our generated kernels. This is due to the
large amount of prior knowledge, experimentation, and engineering effort that
the library authors put into optimizing this operation for this device. By
contrast, our system has only a few heuristics to use when scheduling a kernel. This
demonstrates the continuing value in hand-optimization hot spots, a role which kernel libraries are still well-suited to fill. However, we
find with configurations (b) - (c) that for many shapes which the existing
library has not yet been optimized for (due to less common demand for these
operation sizes), ISAM can produce comparable or slightly better-performing
kernels. Finally, for shapes such as configuration (d) which do not currently
fit the algorithm used in the kernel library well, ISAM can produce
significantly faster kernels.

We note, again, that this benchmark (matrix multiplication) is one which is
quite challenging to our system, since it is already well-optimized by the
kernel library in many cases, and there is no opportunity for ISAM to perform
inter-operation fusion. Nevertheless, we have found that there are situations
in which ISAM can produce performant kernels, which can temporarily be used in
place of the kernel library and as a starting point for hand-optimization.

After comparing ISAM generated kernels against KL kernels, we found that ISAM
would often schedule computations in a way that causes only one processing unit
in a pair to be active, halving our utilization of the chip. These types of
optimizations are well suited to augment the existing heuristics implemented in
Approach classes and remain future work.

\subsubsection{GRU}
\begin{figure}[t!]
  \centering
    \includegraphics[width=1\columnwidth]{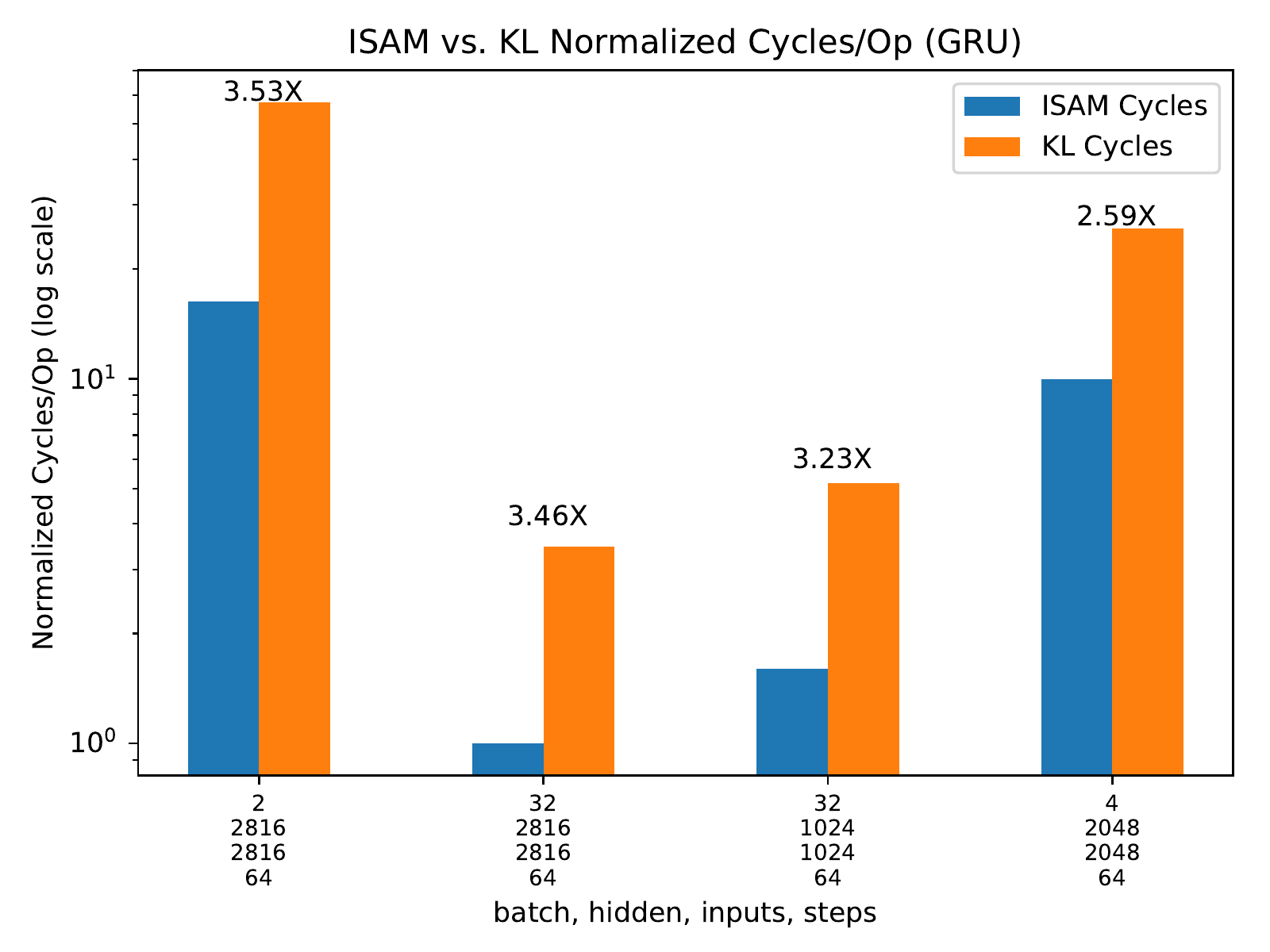}
  \caption{
      A selection of results comparing cycles per operation for ISAM and the
      kernel library (``KL") when executing a GRU cell.
      Although the KL outperforms ISAM for many
      of the underlying operations in the GRU cell (see
      Figure~\ref{fig:matmul_results}), ISAM utilizes
      inter-operation fusion and memory reuse opportunities to
      outperform the equivalent KL computations.
  }
  \label{fig:gru_results}
\end{figure}

To test whether ISAM's potential to benefit recurrent computations, such as
leaving re-used data on register files between kernels or pipelining kernels we
compiled and executed a GRU recurrent neural network over 128 steps, using
matrix shapes adapted from the DeepBench~\cite{deepbench} standardized
benchmark. A selection of our results are presented in
Figure~\ref{fig:gru_results}.

As these results show, ISAM out-performed the composition of kernel-library
operations in all tested GRU cases due to the better fusion of
operations and better intermediate storage that ISAM schedules automatically.

\section{Applicability to General Purpose Architectures}
\label{sec:isamtvm}

As demonstrated above, our methodology can yield very good results on the
chosen case study architecture with matrix instructions. However, the majority
of today's machines running deep learning applications are stock CPUs and GPUs.
Therefore it is extremely valuable to demonstrate if the principles behind ISAM
can be used to reach close to peak performance on those architectures.
Recently, LIBXSMM~\cite{2018arXiv180805567G} demonstrated close to peak
performance on modern x86 CPUs through the use of hand-optimized ``micro-kernels"
(such as 32x32 matrix multiplications) using similar concepts as ISAM
(``dry-run" static scheduling by dynamic programming
optimizations~\cite{DBLP:journals/corr/abs-1802-06905}, ``replay" execution of
optimized instruction streams per PE/core) then manually mapping large
computations to these efficient micro-kernels. However, the reliance on manually-optimized
micro-kernels and hand-written lowering rules makes LIBXSMM similar to the kernel library
on our test architecture.

There are number of ways to use an ISAM-like system to schedule computations on
a traditional, x86 CPU. First, scalar instructions can be represented in ISAMIR
and the system can be run as normal. However, this is difficult for ISAM, as
there are a large number of x86-specific heuristics that would need to be
exposed to ISAM to produce code comparable with LLVM or GCC. Next, methods such
as GEMM from BLAS-like kernel libraries can be exposed as
``pseudo-instructions" to ISAM, allowing ISAM to schedule programs in a similar
way to our test architecture with matrix multiplication instructions while
benefiting from the hand-optimizations in the kernels. We have demonstrated
that this is possible in Section~\ref{sec:results:mapper}. Finally, we have
found that existing compilers such as LLVM can produce extremely performant
output if the input program is reordered to a form that LLVM can correctly
analyze. For example, if the loop nests and buffer dimensions in a convolution
are reordered such that the inner-most loops and most-minor buffer dimensions
are ordered similar to a matrix multiplication, LLVM will automatically
optimize that block of code using device- and algorithm-specific heuristics. In
this way, we can compile programs to x86 devices in a ``full-stack" (but still
efficient) manner, without hand-written kernel libraries.

\begin{figure}[t!]
  \centering
    \includegraphics[width=1\columnwidth]{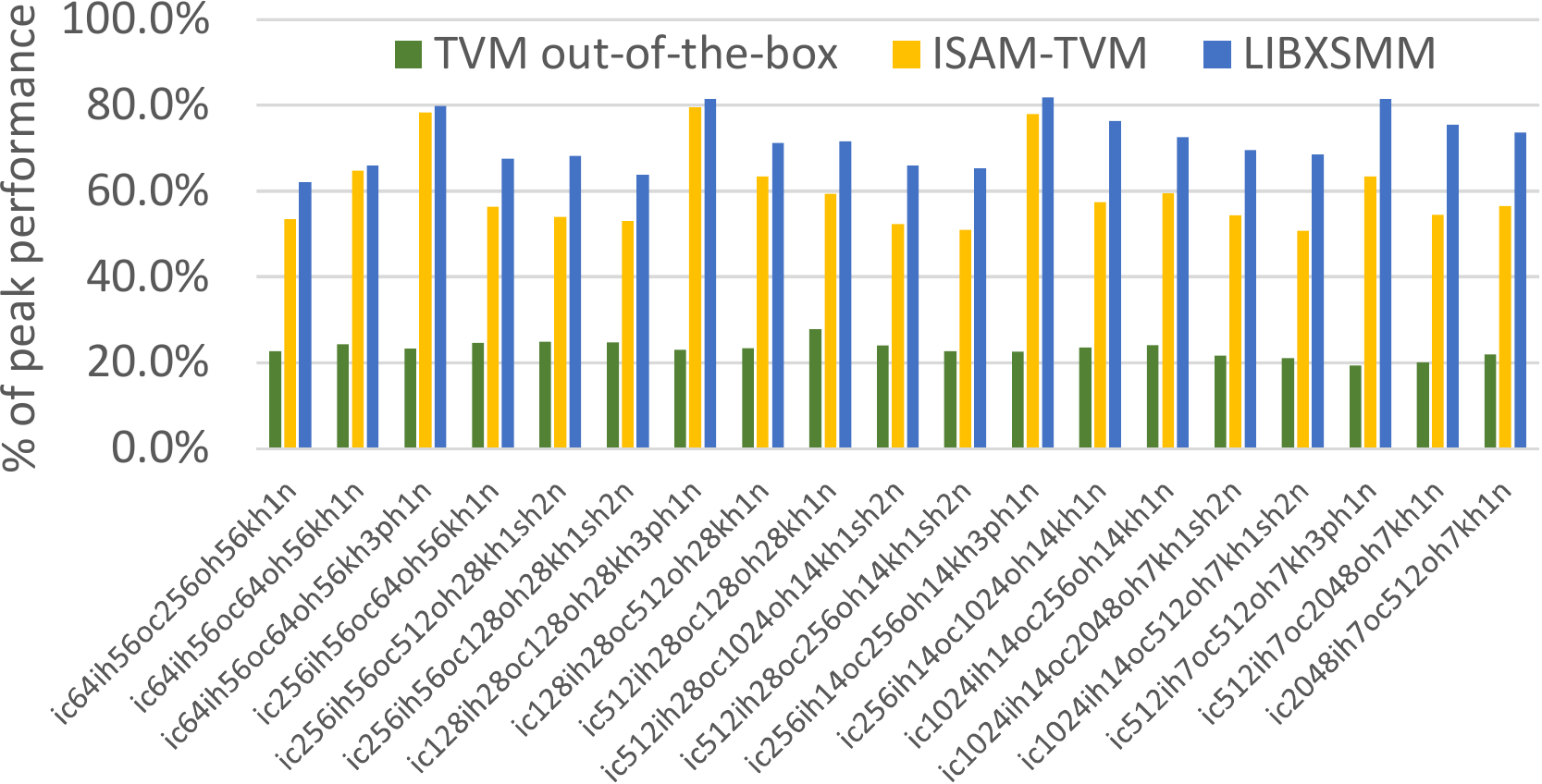}
  \caption{
    Comparing peak performance on an Intel\textregistered{}
    Xeon\textregistered{} Scalable Platinum 8180 processor for TVM, ISAM-TVM
    and LIBXSMM when executing all ResNet-50 layers with the very small
    minibatch 28.
  }
  \label{fig:isamtvm}
\end{figure}

To that end, we can use the mappings found by ISAM to reorder the loop nests
and buffer dimensions in a program in such a way that LLVM can automatically
recognize and optimize such matrix multiplications, then output LLVM IR and
have LLVM perform the final scheduling and compilation. Although, for future
work, we could write an ``LLVM Backend" to do this completely from ISAM (or
expose the LLVM-recognized matrix multiplication as an instruction), we found
that re-using existing work such as TVM would allow us to re-use existing
device- and LLVM-specific heuristics in that system more quickly than rewriting
them for the ISAM system. To that end, following~\cite{2018arXiv180805567G}, we
utilize the mappings that ISAM can find between large convolution and small
matrix multiplications to schedule convolutions in TVM in such a way that LLVM
can correctly optimize the underlying matrix multiplication instructions after
TVM generates and compiles LLVM IR. We refer to this combined system as
``ISAM-TVM."

In order to evaluate the performance of our ISAM-TVM prototype, we run all
inner ResNet-50 layers using TVM out-of-the-box, ISAM-TVM and LIBXSMM on a
single socket Intel\textregistered \, Xeon\textregistered \, Scalable Platinum
8180 CPU with 28 cores at Intel\textregistered{} Advanced Vector Extensions
(Intel\textregistered{} AVX) 512 base-frequency of 1.7\,GHz delivering 3.05 TFLOPS peak performance in
single precision.  Figure~\ref{fig:isamtvm} depicts the achieved performance
and confirms ISAM's general applicability to CNNs using a small matmul approach
for a very small minibatch of 28. Our ISAM-TVM is able to achieve up 85\% of
the KL LIBXSMM (version 1.9-1999) when weighting all layers by their floating
point operations and is able to clearly outperform the default TVM code
generation.  Both ISAM-TVM and LIBXSMM are able to achieve a high fraction of
the 3.05 TFLOPS peak performance for all layers under investigation. The used
memory layout is NCHWc16 for activations and KCRSc16k16
for weights as in~\cite{2018arXiv180805567G}.

\section{Conclusion}
By focusing on a narrower domain of deep learning computation, we were able to
formulate an intermediate representation which is able to encode a broad
variety of common operations and is amenable to efficient pattern matching for
existing and exotic new hardware instruction sets. Then by building this
matcher and a static scheduler over a flexible graph hardware abstraction, we
were able to generate efficient code for two hardware architectures when
compared to more traditional classical kernel library approaches.

\bibliography{cites}{}

\begin{thebibliography}{15}
\providecommand{\natexlab}[1]{#1}
\providecommand{\url}[1]{\texttt{#1}}
\expandafter\ifx\csname urlstyle\endcsname\relax
  \providecommand{\doi}[1]{doi: #1}\else
  \providecommand{\doi}{doi: \begingroup \urlstyle{rm}\Url}\fi

\bibitem[Aho et~al.(1989)Aho, Ganapathi, and Tjiang]{aho1989code}
Aho, A.~V., Ganapathi, M., and Tjiang, S.~W.
\newblock Code generation using tree matching and dynamic programming.
\newblock \emph{ACM Transactions on Programming Languages and Systems
  (TOPLAS)}, 11\penalty0 (4):\penalty0 491--516, 1989.

\bibitem[Baidu(2017)]{deepbench}
Baidu.
\newblock {DeepBench}: Benchmarking deep learning operations on different
  hardware, 2017.
\newblock URL \url{https://github.com/baidu-research/DeepBench}.

\bibitem[Chen et~al.(2018{\natexlab{a}})Chen, Moreau, Jiang, Shen, Yan, Wang,
  Hu, Ceze, Guestrin, and Krishnamurthy]{chen2018tvm}
Chen, T., Moreau, T., Jiang, Z., Shen, H., Yan, E., Wang, L., Hu, Y., Ceze, L.,
  Guestrin, C., and Krishnamurthy, A.
\newblock Tvm: End-to-end optimization stack for deep learning.
\newblock \emph{arXiv preprint arXiv:1802.04799}, 2018{\natexlab{a}}.

\bibitem[Chen et~al.(2018{\natexlab{b}})Chen, Zheng, Yan, Jiang, Moreau, Ceze,
  Guestrin, and Krishnamurthy]{autotvm}
Chen, T., Zheng, L., Yan, E., Jiang, Z., Moreau, T., Ceze, L., Guestrin, C.,
  and Krishnamurthy, A.
\newblock Learning to optimize tensor programs.
\newblock \emph{arXiv preprint arXiv:1805.08166}, 2018{\natexlab{b}}.

\bibitem[{Cordella} et~al.(2004){Cordella}, {Foggia}, {Sansone}, and
  {Vento}]{sgisolver}
{Cordella}, L.~P., {Foggia}, P., {Sansone}, C., and {Vento}, M.
\newblock A (sub)graph isomorphism algorithm for matching large graphs.
\newblock \emph{IEEE Transactions on Pattern Analysis and Machine
  Intelligence}, 26\penalty0 (10):\penalty0 1367--1372, 2004.

\bibitem[Cyphers et~al.(2018)Cyphers, Bansal, Bhiwandiwalla, Bobba, Brookhart,
  Chakraborty, Constable, Convey, Cook, Kanawi, et~al.]{cyphers2018intel}
Cyphers, S., Bansal, A.~K., Bhiwandiwalla, A., Bobba, J., Brookhart, M.,
  Chakraborty, A., Constable, W., Convey, C., Cook, L., Kanawi, O., et~al.
\newblock Intel ngraph: An intermediate representation, compiler, and executor
  for deep learning.
\newblock \emph{arXiv preprint arXiv:1801.08058}, 2018.

\bibitem[Demmel \& Dinh(2018)Demmel and
  Dinh]{DBLP:journals/corr/abs-1802-06905}
Demmel, J. and Dinh, G.
\newblock Communication-optimal convolutional neural nets.
\newblock \emph{CoRR}, abs/1802.06905, 2018.
\newblock URL \url{http://arxiv.org/abs/1802.06905}.

\bibitem[{Georganas} et~al.(2018){Georganas}, {Avancha}, {Banerjee},
  {Kalamkar}, {Henry}, {Pabst}, and {Heinecke}]{2018arXiv180805567G}
{Georganas}, E., {Avancha}, S., {Banerjee}, K., {Kalamkar}, D., {Henry}, G.,
  {Pabst}, H., and {Heinecke}, A.
\newblock {Anatomy Of High-Performance Deep Learning Convolutions On SIMD
  Architectures}.
\newblock \emph{ArXiv e-prints}, August 2018.

\bibitem[Keutzer(1987)]{Keutzer:1987}
Keutzer, K.
\newblock Dagon: Technology binding and local optimization by dag matching.
\newblock In \emph{Proceedings of the 24th ACM/IEEE Design Automation
  Conference}, DAC '87, pp.\  341--347, New York, NY, USA, 1987. ACM.
\newblock ISBN 0-8186-0781-5.
\newblock \doi{10.1145/37888.37940}.
\newblock URL \url{http://doi.acm.org/10.1145/37888.37940}.

\bibitem[Leary \& Wang(2017)Leary and Wang]{tfxla}
Leary, C. and Wang, T.
\newblock {XLA}: Tensorflow, compiled!
\newblock TensorFlow Dev Summit, 2017.

\bibitem[Ne{\v{s}}etril \& de~Mendez(2012)Ne{\v{s}}etril and
  de~Mendez]{nevsetril2012sparsity}
Ne{\v{s}}etril, J. and de~Mendez, P.~O.
\newblock Sparsity: Graphs, structures, and algorithms, volume 28 of algorithms
  and combinatorics, 2012.

\bibitem[Ragan-Kelley et~al.(2013)Ragan-Kelley, Barnes, Adams, Paris, Durand,
  and Amarasinghe]{ragan2013halide}
Ragan-Kelley, J., Barnes, C., Adams, A., Paris, S., Durand, F., and
  Amarasinghe, S.
\newblock Halide: a language and compiler for optimizing parallelism, locality,
  and recomputation in image processing pipelines.
\newblock \emph{ACM SIGPLAN Notices}, 48\penalty0 (6):\penalty0 519--530, 2013.

\bibitem[Sifre \& Mallat(2014)Sifre and Mallat]{sifre2014rigid}
Sifre, L. and Mallat, S.
\newblock \emph{Rigid-motion scattering for image classification}.
\newblock PhD thesis, Citeseer, 2014.

\bibitem[{Vasilache} et~al.(){Vasilache}, {Zinenko}, {Theodoridis}, {Goyal},
  {DeVito}, {Moses}, {Verdoolaege}, {Adams}, and {Cohen}]{tensorcomp}
{Vasilache}, N., {Zinenko}, O., {Theodoridis}, T., {Goyal}, P., {DeVito}, Z.,
  {Moses}, W.~S., {Verdoolaege}, S., {Adams}, A., and {Cohen}, A.
\newblock Tensor comprehensions: Framework-agnostic high-performance machine
  learning abstractions.
\newblock \emph{arXiv preprint arXiv:1802.04730}.

\bibitem[Vertex.ai(2017)]{plaidml}
Vertex.ai.
\newblock Announcing {PlaidML}: Open source deep learning for every platform,
  2017.
\newblock URL \url{http://vertex.ai/blog/announcing-plaidml}.

\end{thebibliography}
\bibliographystyle{sysml2019}

\end{document}